\documentclass[aps,pra,twocolumn,amsmath,amssymb,showpacs,superscriptaddress]{revtex4}
\usepackage{graphicx,epsfig}
\usepackage{times,amsmath,amssymb,latexsym}
\usepackage{rotating}
\usepackage{stmaryrd}

\newcommand{\bs}{\boldsymbol}
\newcommand{\ket}[1]{\left|#1\right\rangle} %|"cosa">
\newtoks\nslashfraction\nslashfraction={.13}\newcommand{\nslash}[1]{\,\setbox0\hbox{$#1$}\setbox0\hbox to\the\nslashfraction\wd0{\hss\box0}/\box0}  %Para poder poner. Letras "slashadas" con el comando \nslash{"letra"}
\begin{document}

\title{Hall response of interacting bosonic atoms in strong gauge fields: from condensed to fractional quantum Hall states.}
\author{H. Pino}
\affiliation{Departament d'Estructura i Constituents de la Mat\`eria, Facultat de F\'isica, Universitat de Barcelona, Diagonal 645, E-08028 Barcelona, Spain}

\author{E. Alba}
\affiliation{Instituto de F\'{i}sica Fundamental, IFF-CSIC, Calle Serrano 113b, 28006 Madrid, Spain.}

\author{J. Taron}
\affiliation{Departament d'Estructura i Constituents de la Mat\`eria, Facultat de F\'isica, Universitat de Barcelona, Diagonal 645, E-08028 Barcelona, Spain}
\affiliation{Institut de Ci\`encies del Cosmos, E-08028 Barcelona, Spain. }

\author{ J. J. Garcia-Ripoll}
\affiliation{Instituto de F\'{i}sica Fundamental, IFF-CSIC, Calle Serrano 113b, 28006 Madrid, Spain.}

\author{N. Barber\'an}
\affiliation{Departament d'Estructura i Constituents de la Mat\`eria, Facultat de F\'isica, Universitat de Barcelona, Diagonal 645, E-08028 Barcelona, Spain}

\begin{abstract}

Interacting bosonic atoms under strong gauge fields undergo a series of phase transitions that take the cloud
from a simple Bose-Einstein condensate all the way to a family of fractional-quantum-Hall-type states [M. Popp,
B. Paredes, and J. I. Cirac, Phys. Rev. A 70, 053612 (2004)]. In this work we demonstrate that the Hall response
of the atoms can be used to locate the phase transitions and characterize the ground state of the many-body
state. Moreover, the same response function reveals within some regions of the parameter space, the structure
of the spectrum and the allowed transitions to excited states. We verify numerically these ideas using exact
diagonalization for a small number of atoms, and provide an experimental protocol to implement the gauge fields
and probe the linear response using a periodically driven optical lattice. Finally, we discuss our theoretical results
in relation to recent experiments with condensates in artificial magnetic fields [ L. J. LeBlanc, K. Jimenez-Garcia,
R. A. Williams, M. C. Beeler, A. R. Perry, W. D. Phillips, and I. B. Spielman, Proc. Natl. Acad. Sci. USA 109,
10811 (2012)] and we analyze the role played by vortex states in the Hall response.

\end{abstract}

\date{\today }
\maketitle

\section{INTRODUCTION}

Within the last few years it has been demonstrated that cold
gases of neutral atoms play an important role as quantum
emulators, as these systems provide controllable devices
available to simulate other systems of interest or theoretical
models. The great flexibility of cold gases for which different
types of atoms or interactions as well as their environment
can be selected nearly at will, offer many possibilities \cite{dal,blo,blo1}.
Among them is the simulation of the quantum Hall effect.

So far two main alternatives to engineer sufficiently strong
artificial gauge fields to generate fractional-quantum-Hall
(FQH) states have been proposed. One is by atom-laser
coupling and the other using rotating potential traps. When
a neutral atom moves in a properly designed laser field,
its center-of-mass motion may mimic the dynamics of a
charged particle in a magnetic field. When the atom follows
adiabatically one of its dressed states (i.e., local eigenstates
of the atom-light coupling), artificial magnetism emerges, due
to the accumulation of the Berry phase \cite{dal,jul}. An alternative
procedure to generate gauge fields is by the rotation of the trap
potential that confines the system \cite{coo,fet}, where the rotation
frequency $\Omega$ plays the role of the magnetic field. In the regime
of high magnetic flux, or rapid rotation in the case of a bosonic
cloud, theory predicts the appearance of strongly correlated
phases. These phases can be viewed as the bosonic version
of FQH states. Moreover, many-particle systems provide a
rich variety of different phases of quantum matter. Different
proposals of classification can be found in literature \cite{sac,wen,che}.

Nowadays it is experimentally feasible to create artificial
fields for ultracold atoms trapped in optical lattices (see \cite{mac},
Chap. 4 where a rather exhaustive analysis is made of the
state of the art). Proposals for Abelian and even non-Abelian
gauge potentials exist, which in some cases are candidates
to exhibit Hall effect under special setups \cite{gol,nie,jak,bha,pal,umu}, and
furthermore, some recent articles refer to valuable schemes
to measure signatures of the Hall effect \cite{alb,gol1,wan,gol2}. A new
generation of proposals to generate FQH states is given
by Refs\cite{coo5,yao}. However, experimental difficulties have
prevented the observation of strongly correlated states. On the
one hand, the use of laser beams comes with some drawbacks,
such as heating of the atoms due to residual spontaneous
emission \cite{dal}, and on the other hand, in the case of rotating
trap, if the number of atoms is large (as is usually the case
in experiments), the $N$-dependent critical rotation frequency
needed to enter the strongly correlated regime is so close to the
trap frequency that the system becomes unstable. Nonetheless,
vortex states have been experimentally obtained using both
techniques \cite{mad,lin}.

In practice, the measurement of interesting observables in
an experiment with few atoms and high density, the regime
in which interactions are most relevant, is very challenging
because we may not have access to the spatial density profile.
With this in mind, we present an experimental proposal to simulate
the few-particle system in each site of an optical lattice,
wherewe choose the rotating trap-potential alternative. Several
interesting features are stressed: First, it is an ideal playground
to test the properties of a Hall system with strong interactions.
Second, the suggested experiment allows the study of a larger
number of atoms than what is computationally feasible with
our techniques. Third, as the window of observability (the
range of $\Omega$ values) of strongly correlated states increases as $N$
decreases, the critical value of $\Omega$ from which FQH states of
few atoms are observable is far from the instability region. Or
in other words, taking advantage of the possibility to have few
atoms per site ($N\le 10$) we solve the experimental difficulty
that has so far prevented the observation of FQH states for
large systems. And finally, we introduce a method to measure
the linear response using time-of-flight images, which do not
require spatial resolution of the atomic profile in each lattice
site.

Measuring characteristic properties of the highly correlated
state is a very relevant problem. We will show below
that the linear response function related to the Hall effect
provides valuable information about the eigenstates and phase
transitions in the strongly correlated atomic system. This
requires a measurement of the population of the ``scissor
mode'' ($\langle x y \rangle$), which is Hall excited by the ``breathing-mode''
perturbation $H_{pert}\propto y^2 \cos(\omega t)$; see \cite{leb} for the first
experimental observation of the Hall effect with atoms in the
mean-field regime. We expect that experimentally the same
informationwill be available in the Hall response of the system,
at least in the weak perturbation limit. We use the rotating
frequency as the driven parameter and find that as the rotation
increases, new phases emerge which are directly related to
angular momentum transitions \cite{wil,coo2,dag1,dag2}. To characterize the
nature of the many-body ground states within a phase, we
analyze the role played by its excitation spectrum in the Hall
dynamical response. We find that the Hall response increases
at phase transition points, and that it is modified in the presence
of vortex states.

This paper is organized as follows: In Sec. II we present
the model of the unperturbed system in two parts. In Sec. IIA
we show the equations used to perform exact diagonalization
of a rotating system to obtain the ground state (GS) and the
excitations. In Sec. IIB we propose an optical lattice setup with
the implementation of rotating traps in each independent site
and infer the expression of the trap frequency used in Sec. IIA
as a function of the experimental parameters. In Sec. III we
display the expressions used within the linear response theory
to obtain the Hall response function at an individual site and
how it can be implemented in the lattice. In Sec. IV we analyze
our results and give an interpretation. Finally in Sec. V we
present our conclusions. In the Appendix, we give a detailed
explanation of our optical lattice proposal.

\section{Model of the unperturbed Hall system}
\subsection{ Analytical background: A single rotating trap}
\label{model}

Inside each site of the optical lattice, we assume a twodimensional
system of $N$ bosonic atoms of mass $M$. The cloud
is trapped in a rotating parabolic potential of frequency $\omega_{\perp}$ and
rotation $\Omega$ along the $z$ axis. In the rotating frame of reference,
the Hamiltonian reads \cite{caz}

\begin{equation}
H_0\,\,\,=\,\,\,H_{sp}+H_{int},
\end{equation}
the single particle (sp) part given by,
\begin{equation}
H_{sp}\,\,\,=\,\,\,\frac{1}{2M}(\bs p+\bs A )^2 + \frac{1}{2}M\left( \omega_{\perp}^2-\frac{(B^*)^2}{4M^2}\right)r^2+W
\end{equation}
with
\begin{equation}
A_x=\frac{B^*}{2}y\,\,\,,\,\,\,A_y=-\frac{B^*}{2}x
\end{equation}

where the particular selection of the symmetric gauge has
been made in the definition of $\bf A$, being $B^*=2M\Omega$ a
constant artificial magnetic field directed downward along the
$z$ direction and $\bs r =(x,y)$. From now on we consider $M=1/2$
and $\hbar=1$ and choose $\,\lambda_{\perp}=\sqrt{\frac{\hbar}{M\omega_{\perp}}}=\sqrt{2/\omega_{\perp}}\,$, $\,\hbar\omega_{\perp}/2\,$ and $\,\omega_{\perp}/2\,$ as units of length, energy, and frequency, respectively.
With our unit of length, $\,\omega_{\perp}=2$. $W$ fixes a term that breaks
the isotropy of the trapping potential and is given by

\begin{equation}
W(x,y)=2\,\tau \,M\,\omega_{\perp}^2\,(x^2-y^2)
\end{equation}
where the dimensionless parameter $\,\tau\,$ measures the strength of the anisotropy.  With this term, the part of the trapping potential which is independent of $B^*$ can be rewritten as

\begin{equation}
\label{eqn:ssitepot}
V_{trap}(x,y)=\frac{M}{2}(\omega_x^2x^2+\omega_y^2y^2)
\end{equation}

where $\omega_x^2=\omega_{\perp}^2(1+4\tau)$ and $\omega_y^2=\omega_{\perp}^2(1-4\tau)$, being  $\tau\leq 1/4$.  For the sake of stability we require $\Omega(=B^*)\leq \omega_{\perp} \sqrt{1-4\tau} = 2\sqrt{1-4\tau}$. 

We model the atomic interaction by a 2D contact potential characterized by,
\begin{equation}
H_{int}=\frac{\hbar^2 g}{M}\sum_{i<j} \delta^{(2)}(\bs r_i-\bs r_j)
\end{equation}
where $g=\sqrt{8\pi}a/\lambda_z$ is the dimensionless coupling, $a$ is the 3D scattering length and $\lambda_z=\sqrt{\hbar/M\omega_z}$. We assume $\omega_z$ the trap frequency in the $z$ direction much larger than any of the energy scales involved, in such a way that only the lowest level is occupied, the dynamic of the system is frozen in the $z$ axis and  can be considered as two dimensional.  

The analytical solutions of the single-particle isotropic problem ($\tau=0$) is given by the Fock-Darwin wave functions \cite{jac}:
\begin{equation}
\phi_{nm}(\theta,r)=\frac{e^{im\theta}}{\sqrt{2\pi}}\sqrt{\frac{2n!}{(m+n)!}}e^{-r^2/2} r^{m}L_n^m(r^2)
\end{equation}
 where $n$ $(=0,1,2,...)$ is the Landau level, $m$ $( m \geq -n )$ is the single-particle angular momentum and $L_n^m$ is the Laguerre polynomial \cite{gra}.

Within the lowest Landau level (LLL) approximation ($n=0$), we choose the set of Fock-Darwin functions given by, $\phi_{0m}(\theta,r)=\frac{e^{im\theta}}{\sqrt{\pi m!}}\,e^{-r^2/2}\, r^m\,$ to represent operators and functions in the second quantized formalism. The single-particle eigenenergies are
\begin{equation}
E_{0m}=\hbar(\omega_{\perp}-B^*)m+\hbar\omega_{\perp}. 
\end{equation}
 
To formulate the many-body problem, we consider the set of many-body Fock states $\ket{n_1,n_2,...}$ where $n_i$ refer to the occupation of the single-particle states $\phi_{0m_i}$. From now on, we will omit the Landau level index $n=0$. We truncate the single-particle state label $m$ where $m_{max}$ and consequently the dimension of the Hilbert space is fixed by the requirement of convergence of the results. We perform exact diagonalization and for convenience, analyze the Hilbert space of many-body Fock states in subspaces with fixed total angular momentum $L$. In general $L$ is a non-well-defined parameter. We want to stress that the convergency condition in much more demanding for the Hall response function than  for the GS, as for the Hall response all the excited states are involved and this is especilly the case if some anisotropy is included. 

\subsection{Optical lattice implementation}

We now propose an optical lattice implementation that produces the rotating trapped system assumed for each site in the previous section. The spirit of the proposal follows the work of Popp et al.~\cite{pop}, using an optical lattice potential to isolate a few atoms per site and simulate the gauge field through a fast rotation per site. Unlike the setup by Gemelke et al.~\cite{gem}, our proposal relies on a optical superlattice with square geometry to create the rotation, not on the controlled interference of multiple beams with a time-averaged triangular lattice potential. 

More precisely, we consider a trapping potential in the $z$ direction stronger than any energy scale in our problem, so that our system can be regarded as purely bidimensional as was pointed out previously. We will assume a lattice potential deep enough to supress tunneling between sites; in other words, each lattice site is to be regarded as an incoherent copy of the same experiment. The presence of the lattice is therefore important not only for confining a few-particles system, but also for signal amplification.

We will now explain how the trapping potential can be implemented by modulations of the laser intensities. The laboratory coordinate frame will be denoted with uppercase letters ($X,Y$), while the rotating frame will be denoted with lowercase ($x,y$), as in Sec.~\ref{model}. These reference frames are related by a 2D-rotation matrix of angle $\Omega t$. The harmonic expansion of each lattice site around its minimum yields the trap potential described in the rotating frame in Eq.~(\ref{eqn:ssitepot}). This potential can be realized by three pairs of laser beams in a standing-wave configuration: two of them with the same wavelength $\lambda$ in the $X$  and $Y$ directions, respectively, and a third one with wavelength $\lambda'=\sqrt{2}\lambda$ in the tilted $X+Y$ direction (i.e., along the line $Y=X$). The laser intensities associated to this configuration are (see the Appendix) 

\begin{eqnarray}
\label{Latticepotential}
I(X,t) &=& I_X(t)\sin^2(kX), \nonumber \\
I(Y,t) &=& I_Y(t)\sin^2(kY), \nonumber  \\
I(X,Y,t) &=& I_{XY}(t)\sin^2(k'(X+Y)),
\end{eqnarray}
where
  
\begin{eqnarray}
\label{Latticepotential}
I_X(t) &=& V_0 [1+4\tau \left( \cos(2 \Omega t)-\sin(2 \Omega t) \right) ], \nonumber \\
 I_Y(t) &=& V_0 [1-4\tau \left( \cos(2 \Omega t)+\sin(2 \Omega t) \right) ],  \nonumber  \\
 I_{XY}(t) &=& V_0\, 8\,\tau \,\sin(2 \Omega t).
\end{eqnarray}
The intensity modulations tune both the anisotropy $\tau$ and the rotation frequency $\Omega$, while preserving the average trap frequency, $\,V_0=M \omega_{\perp}^2 /(2k^2)\,$. In particular, Eq.(19) in the Appendix shows an explicit derivation of the trapping frequency $\omega_{\perp}$ as a function of the parameters that define the lasers building the optical lattice:
\begin{equation}
 \omega_{\perp}^2=\frac{32}{M} Re(\alpha)k^2E_0^2.
\nonumber 
\end{equation}

Figure 1 displays a contour plot of the periodic trapping potential for two different times. Deformation and rotation is shown explicitly. 
\begin{figure}
\label{FigLat}
\includegraphics [width=\linewidth]{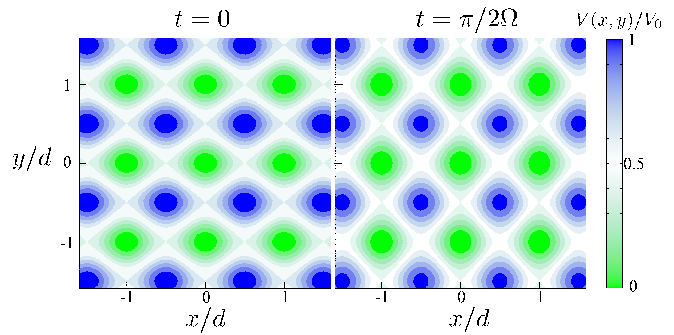}
\caption{Contour plot of the sum of potentials given by the intensities in Eq.9 with $\tau=0.1$ at $t=0$ (left) and $t=\pi/2\Omega$ (right). The figure shows how the resulting potential is a square lattice of anisotropic harmonic oscillators which are rotating individually with frequency $\Omega$. $d=\pi/k$, $k$ being the wavelength of the laser. } 
\end{figure}

The state is prepared by adiabatically loading a BEC into the optical lattice with a fixed small anisotropy $\tau$ and a predefined rotation $\Omega$ for each site. The system will be initially in the approximate ground state, where the chemical potential defines the inhomogeneous atom number density in the central region of the experiment. Once the system is in a ground state with possibly nonzero angular momentum, the lattice site anisotropy $\tau$ is adiabatically switched off~\cite{pop,ron}. Following this procedure, the starting symmetric stationary state per site is prepared and its linear response can then be analyzed.

\section{Linear response function}
 
Once we have the possibility to simulate magnetic fields acting on charged particles in an effective way, in the next step we are going to characterize our interacting many-body states using its susceptibility under weak perturbations. The philosophy is to perturb the system using a small oscillating term that moves the system in one direction (say, the $y$ direction) generating mass current and to measure the response of some observable $A$ that captures the torque experienced by the density distribution due to the presence of an effective magnetic field perpendicular to the $xy$ plane. Significant values of the observable $A$ mean significant ability of the unperturbed system to manifest Hall effect. 
Inspired by a recently published experimental observation of the superfluid Hall effect \cite{leb}, we made the appropriate selection of the observable $A$ and the perturbation.

We calculate the linear Hall response of the sequence of stationary states generated at increasing values of $\Omega$. Our goal is to characterize the many-body states by quantifying their Hall behaviour and analyze the role of the excitation spectrum in the response.
Once the diagonalization is done for a fixed $\Omega$, the eigenstates and eigenvalues of the Hamiltonian are known. Let us call them $\ket{\nu}$ and $E_\nu$, $\nu=0,1,2,..,n_d-1$, respectively, $n_d$ being  the dimension of the Hilbert space considered in the diagonalization.

We choose for the periodic perturbation an extra term in the Hamiltonian $H=H_0+H_{pert}$ given by $\,H_{pert}=M \,\epsilon \,\omega_{\perp}^2\,e^{\eta t} cos(\omega t)B\,$ where $\,B=\sum_i^N y^2_i\,$ and for $\,A=\lambda_{\perp}^{-2}\sum_i^N (xy)_i\,$, $\epsilon$ being a dimensionless small parameter to ensure a perturbation treatment. Had we tried with the operator $B\sim \sum_i^N y_i$ (equivalent to a constant force in the $y$ direction), some nonzero Hall response would also have been obtained \cite{leb}. However we would not expect any contribution from particle interactions, a necessary ingredient in the FQH effect, because this dipolar perturbation produces the center-of-mass displacement of the whole system. However, further analysis is necessary in this direction.

The parameter $\eta$ is assumed to be small enough to ensure adiabatic evolution of the system starting from $t_0\rightarrow -\infty$ when the perturbation is negligible, up to the stationary regime.
From standard linear response theory at zero temperature \cite{pit} we obtain
\begin{equation}
\Delta \langle A \rangle=2 \,\,\epsilon \,\,|\chi(\omega)|\,\, cos(\omega t+\delta)
\end{equation}
where $\Delta \langle A \rangle$ means the change of the expected value of $A$ from the remote past when the perturbation was not active, to the moment when the measurement is performed. $\delta$ is the phase of the complex (dimensionless) $\chi(\omega)$ given by
\begin{eqnarray}
\chi(\omega) & = & \sum_{\nu\ne 0} [ \,\frac{\langle 0|B|\nu\rangle \langle \nu|A|0\rangle}{E_\nu-E_0+\omega+i\eta}
\nonumber
\\
& + & \frac{\langle 0|A|\nu\rangle \langle \nu|B|0\rangle}{E_\nu-E_0-\omega-i\eta}\,\,]\,=\,|\chi(\omega)|e^{i\delta(\omega)}.
\end{eqnarray}
The sum is extended to all excitations. However, only  quadrupolar excitations have non-zero contributions
due to the quadrupolar nature of the perturbation $B$. Being more explicit, in the second quantized formalism the operators $A$ and $B$ take the form
\begin{eqnarray}
\hat{B} & = & -\frac{1}{4}\,\sum_m (\sqrt{m(m-1)}a^{\dag}_m a_{m-2}
\nonumber
\\
& + & \sqrt{(m+1)(m+2)}a^{\dag}_m a_{m+2}-2(m+1)a^{\dag}_m a_m),	
\nonumber
\end{eqnarray}
\begin{eqnarray}
\hat{A} & = & \frac{1}{4i}\,\sum_m (\sqrt{m(m-1)}a^{\dag}_m a_{m-2}\hskip2.6cm
\nonumber
\\
& - & \sqrt{(m+1)(m+2)}a^{\dag}_m a_{m+2}),	
\end{eqnarray}
which can only change the ground-state angular momentum in two units.

 Within our lattice proposal, the small driven perturbation $H_{pert}$ can be implemented by a slight modification of the tunable lattice intensities given by

\begin{eqnarray}
 \delta I_X(t) &=& 2\epsilon V_0 \cos (\omega t) [ \sin^2(\Omega t)+ \sin(2 \Omega t)/2], \nonumber \\
 \delta I_Y(t) &=& 2\epsilon V_0 \cos (\omega t) [ \cos^2(\Omega t)+ \sin(2 \Omega t)/2], \nonumber  \\
 \delta  I_{XY}(t) &=& -2\epsilon V_0  \cos (\omega t) \sin(2 \Omega t),
\end{eqnarray}
which produce the required modulated perturbation, as shown in the Appendix. 
This perturbation is maintained for a time $T$ until the stationary-state regime is achieved. The lattice is then switched off abruptly and a time-of-flight (TOF) image of the system is taken. The fast expansion of the atoms maps the Fourier transform of their wavefunctions to position space. Since many copies of the experiment are performed at the same time, time-of-flight images provide the expectation values of the momentum operators, $\langle p_x p_y\rangle$, which are fourier related to the observable, $\langle x y\rangle$, which we study numerically in the next section. It is important to remember that the TOF measurements take place in the laboratory reference frame. Thus, $\langle p_x p_y \rangle$ will have to be reconstructed from the actual measurements through the relation $\langle p_x p_y \rangle (t)=\langle P_x P_y \rangle (t) \cos(2 \Omega t)/2 + \langle P^2_x- P^2_y \rangle (t) \sin(2 \Omega t)/2$. This can be done by inverting the unitary rotation matrix or via filtering with a frequency $\Omega$.

Finally, we will address the experimental feasibility of our proposal. First we analyze the independent lattice sites approximation. We can set an upper bound to the tunneling parameter as $J<V_0 S(V_0)$ where $S$ is the overlap between the ground-state wavefunctions for neighbouring lattice sites. Let us assume a very deep lattice $V_0 \geq 30 E_R$, where $E_R=\hbar^2/ 2 m \lambda^2$ is the recoil energy, which for $^{39}K$ and $^{87}Rb$ in a $\lambda=800$nm lattice is about $E_R/\hbar \sim 50$kHz. For the ground-state wavefunction with $L=4$, chosen to match most of the results shown in the numerical simulations in section~\ref{numerics}, and considering isotropic lattice sites, we numerically estimate the tunneling parameter  between neighbouring sites to be in the order of a few kHz. This value is just 10 times larger than  when considering a ground-state with $L=0$, which means that $V_0$ can be kept constant at all times. Secondly, we must show that the quadrupolar excitations which can be probed with our Hall response perturbations are lower in energy than the lattice bandgap, so that the single-well experiment approximation is fulfilled. For a deep lattice of $V_0=30E_R$, this bandgap can be estimated as $\Delta_{bg} \sim 10 E_R$ \cite{ger}, or $\Delta_{bg}/\hbar \sim 300$ kHz with the same choice of experimental parameters as before. We can see in Fig.~\ref{spect}, noted below, that in our proposed rotation regime $\Omega = 1.8$ $(1.8\omega_{\perp}/2\,\sim 500$ kHz) the first excited states have an energy of about $E_n \,-\,E_0\sim (0.1-0.3)$ or  $(0.1-0.3)\omega_{\perp}/2 \sim (30-80) $ kHz, so that additional bands will not be significantly populated. It is worth noting that, if $\Omega$ is further increased, and the regime for $L=12$ is reached, the value of $V_0$ must be reviewed.

\section{Numerical analysis}
\label{numerics}

\subsection{Ground and excited states}

\begin{figure}[tbp]
	\centering
		\includegraphics*[width=9.0cm]{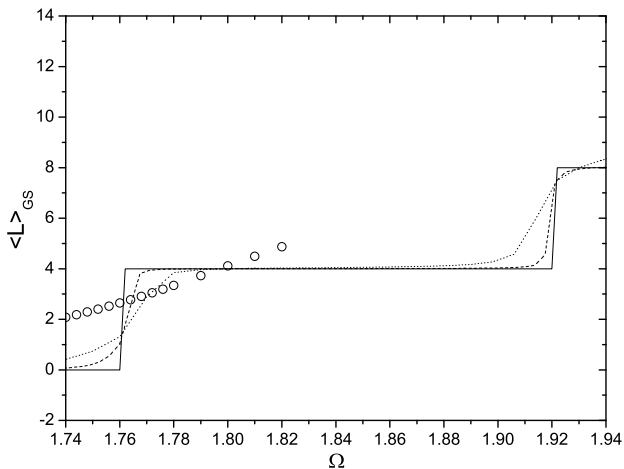}
\caption{Angular momentum of the ground state in units of $\hbar$, as a function of $\Omega$ in units of $\omega_{\perp}/2$ for several anisotropies. $\tau=0$ solid line; $\,\tau=0.8\times 10^{-3}\,$ dashed line; $\,\tau=5.0\times 10^{-3}\,$ dotted line; and $\,\tau=0.025\,$ shown by large dots. The largest possible value for $B^*$ decreases as the anisotropy grows, being $B^*=2M\Omega=1.897$ in our units for $\tau=0.025$}
\label{step1}
\end{figure}

\begin{figure}[tbp]
	\centering
		\includegraphics*[width=9.0cm]{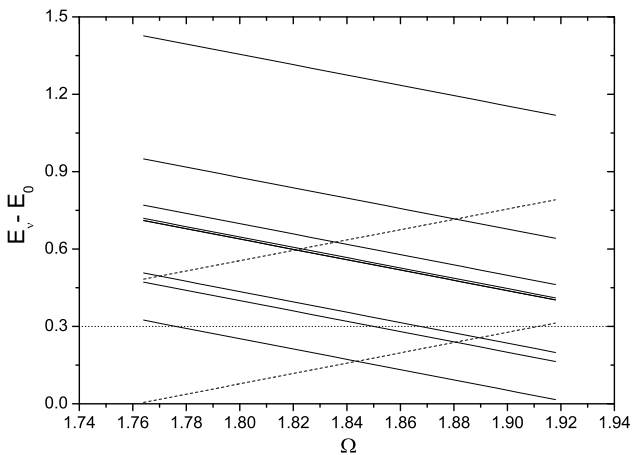}
\caption{Energy gap, $E_\nu-E_0$ versus $\Omega$ for the first step from $\Omega=1.76\,$ to $\Omega=1.92\,$ with $L_{GS}=4$ (see Fig.2). We show only the energies of the quadrupolar excitations, namely, those with $L=6\,$ (solid lines) and $L=2$ (dashed lines). The horizontal dotted line at $0.3$ selects a value for $\omega$ to visualize the crossings at four values of $\Omega$( see Fig.5(b) below). The energies are in units of $\hbar\omega_{\perp}/2$, the frequencies in units of $\omega_{\perp}/2$, and the angular momenta in units of $\hbar$.}
\label{spect}
\end{figure}

From now on we will consider four particles, unless otherwise stated.
Before showing the results of the Hall response function $\chi(\omega)$, it is convenient to have in mind the results in Figs.2 and 3. Figure 2 shows the angular momentum of the GS as a function of $\Omega$ for different anisotropies. For the isotropic case ($\tau=0$, full line), only some special values of the angular momentum are possible GS's, the so-called "magic numbers": $L=0,\,$ is the condensed state (with fidelity $=1\,$, i.e., the overlap between the exact solution and the analytical expression), $L=4\,$ is the Pfaffian state (with fidelity $\,=0.95\,$), $L=8\,$ is the quasiparticle state (with fidelity $\,=0.98\,$) and $L=12\,$ the Laughlin state (with fidelity $\,=1\,$) \cite{pop}. At the values of $\Omega=1.76,\,$ $1.92\,$ and $1.96$, transition jumps (steps) between different angular momenta take place, precisely where several eigenstates with different angular momentum become degenerate. These are the only values of $\Omega$ where anisotropic configurations of the ground-state (e.g., two vortices) are possible as linear combinations of the degenerate states. The analytic expression for the location of the first step is $\,\Omega_c=\omega_{\perp}(1-gN/(8\pi))\,$ \cite{coo2}, where $N$ is the number of particles. Without loss of generality, we assume $gN=3$, for which $\Omega_c= 1.76$. If some anisotropy is considered ($\tau=0.8\times 10^{-3}\,$ or $\tau=5.0\times 10^{-3}\,$ in Fig.2), the steps are smeared out. The transition takes place in a continuous way, over ranges of $\Omega$ of finite size within which anisotropic configurations are possible. If the anisotropy is large ($\,\tau=0.025\,$ in Fig.2) the step structure  disappears and is replaced by an increasing monotonous function, as shown by the large dots. If instead we increase the number of particles maintaining a small anisotropy, the efect on the function $\langle L \rangle_{GS}/\Omega$ is qualitatively similar: The number of steps increases, a sequence of micro-plateaux appears \cite{coo1} and for large values of $\Omega$, $L_{GS}$ becomes a nearly continiuos increasing function similar to the one shown in Fig.2 for large anisotropy.   

Figure 3 displays the spectrum $(E_\nu-E_0)$ versus $\Omega$ over the plateau $L_{GS}=4$ in the isotropic case. Only the quadrupolar excitations to $L=6$ and $L=2$ are considered, since they are the significant ones for our choice of perturbation and observable operators $A$ and $B$ (the operators $\,y^2\,$ and $\,xy\,$ change $L$ in $\pm 2$ units; see Eq.(13)). Once $\omega$ is fixed, every crossing of a constant horizontal line at $\,\omega\,$ with one of the lines of the spectrum is a candidate to be a peak of $\chi(\omega)/\Omega\,$ where, the resonant condition cancels the denominator in the second term of the right hand side of Eq(12). For example, if $\,\omega=0.3\,$, as shown in Fig.3 there are four crossings between the horizontal line at $\,0.3\,$ and the excitations with $\,L=6\,$ (with negative slope) or $\,L=2\,$ (positive slope), as shown in Fig.5(b), or for $\,\omega=0.4\,$ there are two crossings, as shown in Fig.6.  

\subsection{The Hall response}

\begin{figure}[tbp]
	\centering
		\includegraphics*[width=8.5cm]{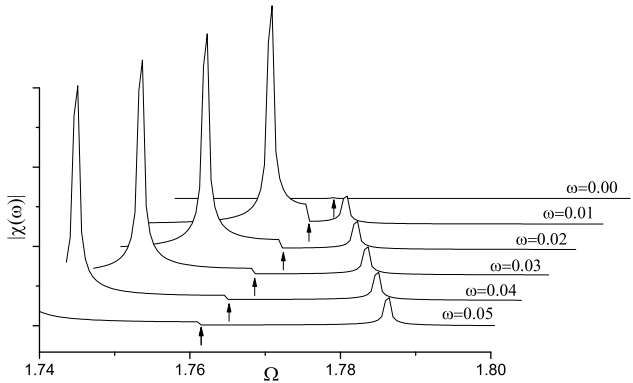}
\caption{Modulus of the Hall response $\chi(\omega)$ (dimensionless) for different values of $\omega$. The tic-labels of the $x$-axis correspond to the lowest curve, the rest are shifted for clarity. The $y$-axis is also shifted. The arrows marc the first step at fixed $\Omega_c$ ($\,=1.76127$), independent of $\omega$. The frequencies are in units of $\omega_{\perp}/2$.}
\label{Hall-resp1}
\end{figure}

\begin{figure}[tbp]
	\centering
		\includegraphics*[width=9.0cm]{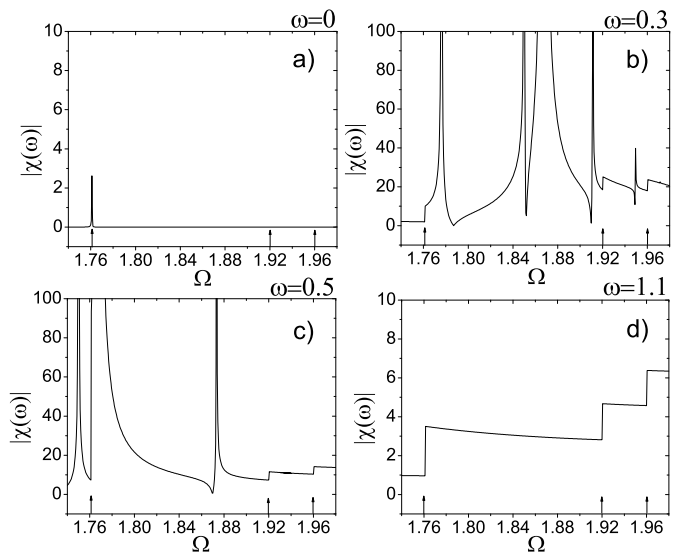}
\caption{Modulus of the Hall response $\chi(\omega)$ (dimensionless) for different values of $\omega$. From (a) to (d) $\omega=0.0,\,$ $0.3,\,$ $0.5,\,$ $1.1\,$. The whole range of $\Omega$ values from $1.74$, where the lowest Landau approximation is still valid, to $1.98$ close to the upper limit at $2$ is considered. The frequencies are in units of $\omega_{\perp}/2$.}
\label{Hall-resp2}
\end{figure}

For the symmetric case, Fig.4 shows a detailed analysis of the structure of $|\chi(\omega)|$ close to the first step at $\Omega_c$ as $\omega$ grows from zero. The small peak (visible in Fig.5(a) for $\omega=0$ in a different scale) splits into two peaks which separate from each other. The important feature is that the height of the step at $\Omega_c$, not related to any resonant condition, also depends on $\omega$. The small peak moves to the right following the crossings between the positive-slope line of the lowest excitation with $L=2$ (see Fig.3) and the values of $\omega$. In Fig.5 four typical cases are shown for the whole range of $\Omega$ starting at $\,1.74\,$. Two different scales are considered: Figs.5(a) and 5(d) up to $10$ and Figs. 5(b) and 5(c) up to $100$. The first plateau from $\,\Omega=1.76\,$ to $\,1.92\,$ is fully dominated by the dynamical response peaks. In contrast, the dynamical response decreases for larger $L$. For $L_{GS}>4$, the peaks disappear beyond $\,\omega=0.4\,$. For $\,\omega=1.1\,$, close to the driving frequency chosen in Ref.\cite{leb}, the peaks completely disappear.

Two comments related to Fig.5 are in order: The physical expected behaviour of the response is given by finit values in the limit $\,\eta\rightarrow 0\,$ (see Eq.(12)). In our simple model, this is not the case as $\,|\chi(\omega)|\,$ diverges in this limit at the resonant points. The inconvenience comes from the fact that we are not considering the width of the excitations that would prevent the divergence. The alternative procedure followed here is to consider a cutoff of $\,\Omega\,$ such that we are close but out of resonance and the results are $\,\eta\,$ independent. Therefore we obtain information about the presence of a peak but not of the details close to its center, where non-linear response is expected due to absorption. At the end, we characterize the ground state.

The other comment refers to the absence of peaks for values of $\,\Omega\,$ and $\,\omega\,$ where there is a crossing ($E_{\nu}-E_0=\omega\,$  (see Fig.3)). The explanation of this possibility is, however, much technical. The absence of a peak means that $\,<0|A|\nu>\,$ and/or $\,<0|B|\nu>\,$ cancels, even though $\,|\nu>\,$ is a quadrupolar excitation. Every eigenstate at $\,\Omega\,$ has associated a specific set of non-zero single-particle occupations, say $\,\{\beta_i\}\,$, where $\,\beta_i\,$ must not be an integer number. Here $i$ labels the single-particle Fock-Darwin function $\,\phi_{m_{i}}\,$ defined previously. If the set defining $\,A|0>\,$ (or $\,B|0>\,$) has no coincidences with the set defining the excitation, then the numerator in Eq.(12) cancels. 

Furthermore, from the results shown in Fig.5 we obtain the following useful information in the line of the characterization of the strongly correlated states. As $\,\Omega\,$ increases, the entanglement of the GS grows since the sp occupations $\,\beta_i\,$ equalize, producing stationary states with internal structure far from condensation  and mean-field description. The extreme case is the Laughlin state with a nearly equal distribution of occupations, even for finite systems, showing large entanglement \cite{jul}. This large entanglement has two consequencies: The Hall response is large (see Fig.5(d)) and the GS is protected against absorptions at odds with the expected behaviour as there is a large amount of possibilities detected as crossings in the spectrum (see Fig.3).
   
Figure 6 displays the case $\,\omega=0.4\,$ with a slight anisotropy ($\,\tau=0.8\times 10^{-3}\,$). As expected, only the steps are modified due to the lifting of the degeneracy produced by the anisotropy. However, the peaks remain unchanged.

As previously noted in Sec. III, the guideline of our performance has close connection with the first experimental observation of the Hall effect with atoms \cite{leb}. A brief explanation of the main ingredients of the experiment is as follows. The initial state is a large ($\,N\sim 10^5\,$) strongly deformed  cloud of bosons in a superfluid regime. By atom-laser coupling, they submit the system to an artificial magnetic field $\,\tilde{B}\,$ perpendicular to the cloud. Next the system is perturbed along the $x$ direction with a time-dependent modification of the trap potential given by $\,\delta U \sim x^2 cos(\omega t)\,$ equivalent to a force linear in $x$. Finally, they measure the time evolution of the second-order moment $\,<xy>\,$ of the density. Their main result is the oscillation of $\,<xy>/t\,$ if $\,\tilde{B}\neq 0\,$ or zero otherwise. This $\,\tilde{B}\,$-dependent correlation  transport (in the $x$ and $y$ directions) is the Hall response. In their Fig.4, the one with which we contrast our results, they show the amplitude of the oscillations of $\,<xy>\,$ as a function of  $\,\tilde{B}\,$
for a fixed $\,\omega\,$. The result is a monotonous increasing function that closely follows the superfluid hydrodynamic equations up to a point where $\,\tilde{B}\,$ is strong enough to generate vortex states; from this point, experimental points depart from the hydrodynamic prediction, and Hall response has lower values than those predicted by the model.

According to our results, for small systems and negligible deformation, classifying the states by their ability to manifest the Hall effect is equivalent to classifying the states by their angular momentum $L$, that is, the phase transitions lie at the steps. Furthermore, following this similarity between Hall response and angular momentum we are naturally brought to an extrapolation:  If we increase $N$ (increasing the number of steps in $L_{GS}/\Omega$ \cite{coo1}) and simultaneously add some deformation, we expect for a Hall response a monotonous increasing function as the one shown by the large dots of Fig.2, in close agreement with Ref.\cite{leb} as discussed in the above paragraph.

To complete our comparison, we analyzed the role played by the vortex states for large values of $\,\tilde{B}\,$. To this end, we moved to $N=6$ and analyzed this posibility.
The sequence of $L_{GS}$ for $6$ atoms is: $\,L=\,0,6,10,12,15,20,24,$ and $30$. In the step produced by the change of $L$ from $10$ to $12$ the creation of a two-vortex state is possible. A very small anisotropy is enough to mix states of different angular momentum and facilitate the precise numerical calculation of the value of $\Omega$ at the vortex state. Different Hall response, related to the general tendency, was expected for a vortex state  due to its different set of sp occupations $\,\{\beta_i\}\,$ as compared with a state laying on a plateau. This difference is related to the more demanding convergency condition in the numerical calculation for the vortex case. For a system laying on a plateau, a single occupation is dominant (the degree of condensation of the state is high) but in contrast two of the occupations play an important role in the vortex state \cite{dag2}. Unexpectedly, our result and the experiment goes in the opposite directions; Fig.7 shows a slight increase. It is difficult to follow the numerics to infer any difference when increasing $N$, or more importantly, when increasing the deformation (up to a quasi-one-dimensional system which is the case in the experiment), two possible reasons of the discrepancy.  

Finally,  a comment about the term ``Hall response''. We have followed the nomenclature used in the experimental work \cite{leb} which is an attempt to mimic the behavoiur of real charges under magnetic and electric fields. We believe that the observation of the torque of the density when the system is displaced in one direction,  which has a significant value only in the case of non-zero $B^*$, is related to the Hall effect and can be used to clasify the states. The absence of response would clasify the state outside the set of $QH$-type states. 
However, it must be noted that it is far from the simulation of the appropriate transport equation given by $\,j_x=\sigma_{xy} E_y\,$ where the conductivity $\,\sigma\,$ or its inverse, the resistivity, show the Hall effect characterized by plateaux when analyzed as a function of the magnetic
 field.

\begin{figure}[tbp]
	\centering
		\includegraphics*[width=8.5cm]{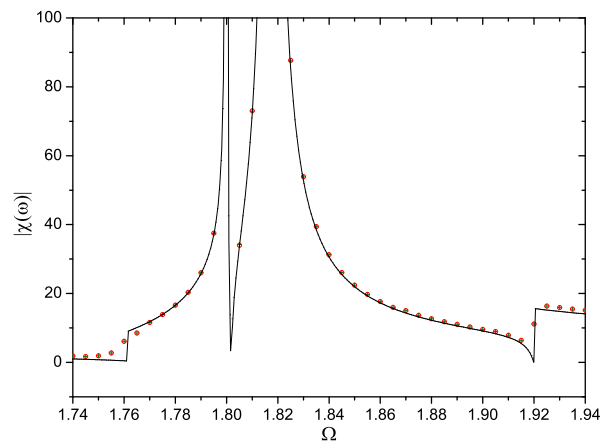}
\caption{Modulus of the Hall response $\chi(\omega)$ (dimensionless) versus $\Omega$ for $\omega=0.4$ with a slight anisotropy ($\tau=0.8\times 10^{-3}$) has been considered. The frequencies are in units of $\omega_{\perp}/2$.}
\label{Hall-resp2}
\end{figure}

\begin{figure}[tbp]
	\centering
		\includegraphics*[width=9.5cm]{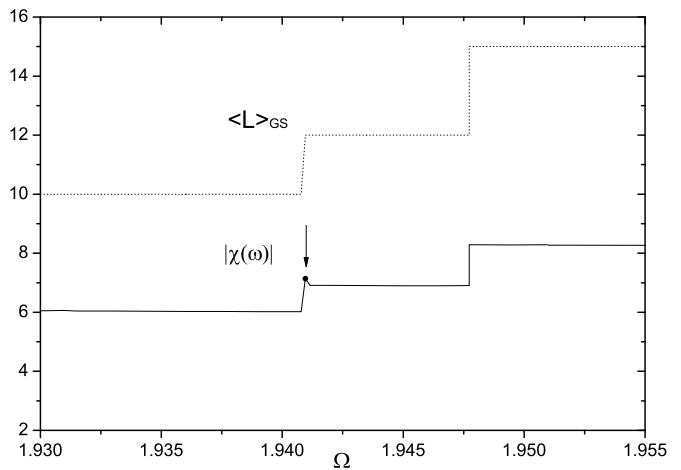}
\caption{Expected value of the angular momentum (in units of $\hbar$, upper part) and the modulus of the Hall response $\chi(\omega)$ (dimensionless, lower part) as functions of $\Omega$ for $N=6$. The upward peak at $\Omega=1.941$ (signaled by an arrow) corresponds to a two-vortex state; $\omega=1.1$ is considered. The frequencies are in units of $\omega_{\perp}/2$.}
\end{figure}

\section{Summary and discussion}

In this work we have studied a system of interacting bosonic atoms under a strong effective magnetic field. Such systems are known to exhibit a variety of phases, ranging from Bose-Einstein condensation to highly correlated states (Pfaffian states, Laughling liquids, etc). While progress in experimental manipulation of ultracold atoms is advancing steadily, there is a huge need for new tools to probe and learn about the physics of strongly correlated states. Our work shows that the linear response can be a very useful method to get information about many-body atomic systems, both from the point of view of phase transitions, signalling the changes between different symmetries (the transition from $L=0$ to $L=N$ e.g., involves a broken parity symmetry at the single particle level \cite{dag2}), and also for accurately analyzing the role played by the excitations. In what follows, we first summarize the whole experimental protocol, next briefly explain our results and finally, present our interpretations and comments. 

We propose an experimental setup using an optical lattice, where several incoherent copies of a few-particle Hall state can be prepared. We have shown that the system can be prepared in a ground state with fixed angular momentum: A BEC is adiabatically loaded into an optical lattice with local deformation and rotation and once the system is in its ground state, the deformation is adiabaticaly switched off while keeping $L$ constant. Next the system is perturbed by using laser intensity modulations at a particular frequency $\omega$, and when the stationary state is achieved, the lattice is switched off and a time-off-flight imaging of the cloud is performed. This measurement provides the density distribution required to build the density moment $<xy>$ (see Eq.(11)) and from it, the linear Hall response $\chi(\omega)$.    

Our results can be summarized as follows: In the isotropic case, the phase transitions related to the Hall response take place at critical values of $\Omega$ where changes in $L$ occur in a stepwise well-known variation. Within a fixed phase, peaks at some specific values of $\Omega$ and $\omega$ provide information about the excited states and characterize the dynamical response of the ground state within this phase. More importantly, we find a relationship between the Hall response and the correlation in the GS. As the correlation (or entanglement) increases, the Hall response grows; in contrast, the dynamical response explicited by the resonant absorption peaks nearly disappears for perturbation frequencies comparable to the excitation energies. For a vortex state of six atoms, we obtain a slight increase of the Hall response as compared with states without vortices at close $\Omega$.

Signatures of Hall response would be obtained if a stepwise structure of $|\chi|$ is observed for large $\omega$ (see Fig.5(d)), each of the plateaux related with a correlated ground state. One of the advantages of our implementation having isolated sites is that there is no influence of the lattice on the structure of the GS as is the case of other possibilities proposed to create FQH states, where a compromise must be achieved: The magnetic flux piercing each cell must be strong enough to produce the FQH state and at the same time, small enough to cancel the influence of the lattice avoiding the  modification of the state \cite{haf}. The identification of the correlated states could be possibly complemented by a local measurement of the two-body correlation function. Other ways to see the correlations between the particles has been proposed in Ref.\cite{sor,haf} using Bragg spectroscopy.      

The presence of an excited state is expected to produce a large response at a particular value of $\omega$ and consequently, energeticaly determined.
In the case of vortex states, our result, which is at odds with the behaviour shown in \cite{leb}, is not conclusive. Extra analysis therefore is necessary for a larger number of atoms and/or vortices. However, having different Hall response as compared with the states without vortices, its presence can be detected experimentaly. Furthermore, according to our analysis, the  stepwise variation of the Hall response, can be used to infer the angular momentum of the isotropic initial ground state.

Two extensions of our analysis could, in principle, be obtained from experimental results: One is the inspection of the response close to the absorption peaks, inaccesible to our linear response calculations. The other is the analysis of the evolution of the Hall response as $N$ is simultaneously increased in each site, as a test of the expected extrapolated results. More importantly, the experimental difficulty that so far has prevented the observation of FQH states for large systems is expected to be solved in our setup, since lower rotation frequencies far from instability at $\,\omega_{\perp}\,$ are required to reach the FQH regime. Furthermore, from our optical lattice implementation, we obtain an analytical expression for the trap potential frequency $\,\omega_{\perp}\,$ as a function of the laser beams and the atoms involved in the experiment.   

The ideas in this manuscript are intimately related to the work by LeBlanc et al~\cite{leb}. However, we must point out that the experimental results reported in that paper were obtained with strongly deformed (with a nearly cigar shape) Bose-Einstein condensates in the hydrodynamic regimes, with a large number of particles and a purely mean-field treatment. On the contrary, the states that we have studied in this work involve a small number of atoms in a symmetric trap, with a strong dominance of the interaction for large rotation speeds. Moreover, we have resolved the transition from vortex-free regime to the regime with vortices, going beyond the hydrodynamical analysis present in that work.

We are indebted with the referee and with M. Rizzi for all their suggestions and comments.
H.P. and N.B. are greatly benefited from discussions with Maciej Lewenstein. This work is partially sopported by the Spanish MEC through the FPU grant No.AP 2009-1761, the EU through the PROMISCE project, the Spanish MINCIN FIS2010-16185, the Consolider CPAN project CSD2007-00042 and the Generalitat de Catalunya Program under contracts 2009SGR502 and 2009SGR21.

\section{Appendix}

Our goal in this Appendix is to obtain the relationship between the Hamiltonian used in Section IIA for a single site (in the rotating frame of reference) and the appropriate laser beam implementation in the laboratory frame to reproduce it. To start, we rewrite Eq.(2) as
\begin{equation}
H_{sp}\,\,\,=\,\,\,\frac{\bf p ^2}{2M}+ V_{trap}-\Omega \hat{L},
\end{equation}
where $V_{trap}$ is given by Eq.(5) and $\hat{L}$ is the angular momentum operator in the $z$ direction. From this expression, it is clear that the only term that must be translated from the rotating to the laboratory frame is $V_{trap}$ since $\frac{\bf p ^2}{2M}\,$ is invariant \cite{lan}. To this end, we analyze Eq.(5) in two terms:

\begin{eqnarray}
V_{trap}(x,y) & = & \frac{M}{2}(\omega_x^2 x^2+\omega_y^2y^2)
\nonumber
\\
& = & \frac{1}{2}M\omega_{\perp}^2(x^2+y^2)+2\tau M\omega_{\perp}^2(x^2-y^2)
\nonumber
\\
& \equiv & V_1+V_2
\end{eqnarray}
or following the convention given in the text about upper and lowercase letters, in the laboratory frame the trap potential reads
\begin{equation}
V_1(X,Y,t)=V_0 k^2(X^2+Y^2),\hskip2cm
\nonumber
\end{equation}
\begin{equation}
V_2(X,Y,t) = 2\tau M\omega_{\perp}^2\,\times \hskip2.8cm
\nonumber 
\end{equation}
\begin{equation}
\hskip1.5cm [X^2 \cos(2\theta)-Y^2 \cos(2\theta)+2XY \sin(2\theta)],
\end{equation}
where we have used the rotation matrix 
\[
\left(\begin{array}{c} x \\ y\end{array}
\right)=
\left(
\begin{array}{ll} \cos(\theta) & \sin(\theta)\\
-\sin(\theta) &  \cos(\theta)\\  \end{array}
\right)
\left(\begin{array}{c} X \\Y \end{array}
\right)
\]
with $\,\theta=\Omega t\,$ and $\,V_0\equiv M\omega_{\perp}^2/(2k^2)\,$.

We now consider on the one hand, that the intensity of a standing wave in the $\,X\,$,$\,Y\,$ and $\,X+Y\,$ directions are given by $\,I=8E_0^2 \sin^2(kX)\,$, $\,I=8E_0^2 \sin^2(kY)\,$ and $\,I=8E_0^2 \sin^2(k'(X+Y))\,$ respectively (being $k'=k/\sqrt{2}$), where $\,E_0\,$ is the amplitud of the electric and $\,k\,$ ($\,k'\,$) is its wave vector. And on the other hand, that the coupling of the laser with the atomic induced dipolar moment $\,\vec{d}\,$ is $\,\vec{d}\cdot \vec{E}= 2 Re(\alpha)I(\vec{r})\,$ where $I$ is the total intensity \cite{mac}. Next, considering these two results, we can express Eqs.(A3) in terms of the laser-atom coupling in a periodic configuration. Previously, we want to stress that from Eqs.(A3) it can be inferred that considering only the first term in the expansion of $\sin^2(kX)$, two standing waves on the $X$ and $Y$ directions are sufficient to generate a symmetric time-independent trap potential, however, a third laser in the $X+Y$ direction is necessary to deform and rotate it. We can rewrite the Eqs.(A3) as 

\begin{equation}
V_1(X,Y,t) \sim V_0(\sin^2(kX)+\sin^2(kY))
\nonumber
\end{equation}
\begin{eqnarray}
V_2(X,Y,t) & \sim & 4 \tau V_0 [\sin^2(kX)(\cos(2\theta)-\sin(2\theta))
\nonumber
\\
\hskip 2cm &  & -\sin^2(kY)(\sin(2\theta)+\cos(2\theta))
\nonumber
\\
\hskip2cm &  & +2\sin^2(k'(X+Y))\sin(2\theta)]\,.
\end{eqnarray}

We assumed that the atomic polarizability is $\alpha_{ij}=\alpha\delta_{ij}$ and made the rotating wave approximation. 
Identifying terms, finally we obtain the main result
\begin{equation}
 \omega_{\perp}^2=\frac{32}{M} Re(\alpha)k^2E_0^2
\end{equation}
 or in other words, we have explicitly obtained the relation between the trapping frequency and the experimental parameters of our configuration (the atomic polarizability $\alpha$, the wave vector $k$, and the intensity of the lasers $E^2_0$). Finally, the term of the Hamiltonian that generates a periodic $\tilde{V}_{trap}$ is given by 
\begin{eqnarray}
\tilde{V}_{trap} & =& I_X(t)\sin^2(kX)+I_Y(t)\sin^2(kY)
\nonumber
\\
&  & +I_{XY}(t)\sin^2(k'(X+Y)),
\end{eqnarray}
where
\begin{eqnarray}
\label{Latticepotential}
I_X(t) &=& V_0 [1+4\tau \left( \cos(2 \Omega t)-\sin(2 \Omega t) \right) ], \nonumber \\
 I_Y(t) &=& V_0 [1-4\tau \left( \cos(2 \Omega t)+\sin(2 \Omega t) \right) ],  \nonumber  \\
 I_{XY}(t) &=& V_0 \,8\,\tau \sin(2 \Omega t).
\end{eqnarray}

Similarly, the expression 
\begin{equation}
H_{pert}(y)= \epsilon M \omega_{\perp}^2\cos(\omega t)y^2,\,\,,
\end{equation}
transformed to the laboratory frame reads
\begin{eqnarray}
H_{pert}(X,Y,t) & = & \epsilon M \omega_{\perp}^2\cos(\omega t)\,\times
\nonumber
\\
&  & (X^2\sin^2(\theta)+Y^2\cos^2(\theta)-2XY\sin(\theta)\cos(\theta))
\nonumber
\\
& \sim & 2 \epsilon V_0 \cos(\omega t)[(\sin^2(\theta)+\frac{1}{2}\sin(2\theta))\sin^2(kX)
\nonumber
\\
\hskip1cm &  & +(\cos^2(\theta)+\frac{1}{2}\sin(2\theta))\sin^2(kY)
\nonumber
\\
\hskip1cm &  & -\sin(2\theta)\sin^2(k'(X+Y))].
\end{eqnarray}
In all the expressions, we considered only the first term in the expansion of $\sin^2(kX)$.

\end{document}